\documentclass[aps,prl,a4paper,twocolumn,superscriptaddress,showpacs]{revtex4}
\usepackage{graphicx,graphics,color,epsfig}
\usepackage[hang]{footmisc}
\usepackage{amsmath}
\usepackage{amsfonts}
\usepackage{upquote}

\usepackage{amssymb}
\usepackage{bm}
\begin{document}
\title{Finite-size scaling of entanglement entropy at the Anderson transition with interactions}
\author{Rui-Lin Chu}
\affiliation{Department of Physics, The University of Hong Kong, Pokfulam Road, Hong Kong, China}
\affiliation{Department of Physics, The University of Texas at Dallas, Richardson, Texas, 75080}
\author{An Zhao}\email{yayayayazam@gmail.com}
\affiliation{Department of Physics, The University of Hong Kong, Pokfulam Road, Hong Kong, China}
\author{Shun-Qing Shen}
\affiliation{Department of Physics, The University of Hong Kong, Pokfulam Road, Hong Kong, China}

\begin{abstract}
 We study the Anderson transition with interactions in one dimension from the perspective of quantum entanglement. Extensive numerical calculations of the entanglement entropy (EE) of the systems are carried out through the density matrix renormalization group (DMRG) algorithm. We demonstrate that the EE can be used for the finite-size scaling(FSS) to characterize the Anderson transition in both non-interacting and interacting systems. From the FSS analysis we can obtain a precise estimate of the critical parameters of the transition. The method can be applied to various one-dimensional models, either interacting or non-interacting, to quantitatively characterize the Anderson transitions.
\end{abstract}

\date{\today}
\maketitle
 In the theory of Anderson localization, all states are localized by disorder in low dimensions when interactions are absent.\cite{Anderson,AALR,PLee_RMP} Theories indeed predict that strong attractive interactions are able to induce a delocalized phase.\cite{Rice_82PRB,Schulz_88PRB} Despite that various widely accepted numerical models of Anderson transitions have neglected interactions, recent studies do suggest that inclusion of interactions is necessary when explaining the experiments.\cite{Slevin_QHE} However, since including the interactions makes the model a many-body system, numerical investigations are computationally challenging. There has been only very limited success in numerical simulations of Anderson transitions with interactions.\cite{PS_98PRL,Cater_05PRB,Schuster_02PRB} Although numerical methods for studying Anderson transitions of single particle systems have developed into standard procedures,\cite{MacKinnon,Huckestein,Slevin_QHE,Slevin_QSH} alternative methods must be sought when interactions are involved since only very limited system sizes can be numerically calculated. Schmitteckert \emph{et al} and Schuster\emph{et al} investigated the system's phase sensitivity of the ground state energy to roughly estimate the delocalized phase,\cite{PS_98PRL,Schuster_02PRB} while Cater and Mackinnon tried to calculate the localization length directly with the transfer-matrix method.\cite{Cater_05PRB}

 For the single particle systems, the finite-size scaling(FSS) method provides the most precise quantitative characterization of the Anderson transition.\cite{Slevin_QHE,Slevin_QSH} A handful of physical quantities can be used for FSS, such as localization length, conductance, density of states, and topological numbers, etc.\cite{Huckestein} However, for the many-body systems, such quantitative characterization of the Anderson transition has been lacking.

 In this Letter we demonstrate that the quantum entanglement entropy (EE) can be adopted as a FSS quantity for Anderson transitions in one dimension. By carrying out the scaling analysis in the framework of one-parameter scaling theory, we give an estimate for various critical parameters of the transition. Before applied to the interacting system, the method is first tested in the non-interacting system, where the result is well consistent with that given by the well established transfer-matrix method.

 The EE is a measure of the quantum correlations in a system.\cite{EE_RMP} For a pure state $|\Psi\rangle$ of bipartite system AB, the EE (von Neumann entropy) is defined as
 \begin{equation}\label{EEdef}
   S=-\mathrm{Tr}\rho_A\log\rho_A=-\mathrm{Tr}\rho_B\log\rho_B,
  \end{equation}
where $\rho_{A(B)}=\mathrm{Tr}_{B(A)}|\Psi\rangle\langle\Psi|$ is the reduced density matrix of subsystem A(B). Studies have shown that it can be used to characterize both quantum criticality and topological phases in a variety of quantum many-body systems.\cite{Vidal_03PRL,Cardy_JSM,Ryu_06PRB,Refael_04PRL,Pollman_10NJP,Turner_11PRB,Luca}
Recently there has also been increasing interest in characterizing the disordered quantum systems through the EE.\cite{Fagotti_11PRB,Igloi_08JSM, Jia_08PRB, Bardarson_12PRL,Laflorencie,Santachiara}

 Since the amount of correlations is restricted by the system's correlation length, localization is naturally manifested in the EE. Consider a one-dimensional(1D) gapless system of length $L$ and partitioned into two halves, when $L\rightarrow \infty$, the EE shows logarithmic divergence in the clean limit as well as in the delocalized phase
 \begin{equation}
 S_H\sim \frac{c}{6} \mathrm{log} L , \nonumber
 \end{equation}
 where c is a universal constant given by the central charge of the associated
conformal field theory. However, when the system is localized the EE only saturates to
\begin{equation}
  S_H\sim \frac{c}{6} \mathrm{log} \xi, \nonumber
\end{equation}
 where $\xi$ is the localization length.\cite{Cardy_JSM} For the strongly localized finite systems in which $\xi$ is well less than $L$, it is possible to extract $\xi$ from the EE's saturation behavior itself.\cite{Berkovits_12PRL} In this Letter, since our main focus is at the critical regime of the Anderson transition where $\xi$ is divergent and hence much larger than the system size $L$, such saturation behavior is absent.

We start by describing our model Hamiltonian, which is a 1D spinless fermion wire of length L with attractive nearest neighbor(NN) interactions
\begin{equation}\label{eq000}
\begin{aligned}
\mathcal{H}=&- t\sum_{i=1}^{L-1}\left(c_i^\dagger c_{i+1}+h.c.\right)+ \sum_{i=1}^{L} \epsilon_i c_i^\dagger c_i
\\ + &\,U\sum_{i=1}^{L-1}\left(n_i-1/2\right)\left(n_{i+1}-1/2\right)
-\mu\sum_{i=1}^{L} c_i^\dagger c_i,
\end{aligned}
\end{equation}
where $c_i$ and $c_i^\dagger$ are the annihilation and creation
operators, $n_i = c_i^\dagger c_i$. The first two terms
constitute the standard Anderson model while the third term
represents the attractive NN interaction ($U < 0$). The disorder is
introduced as random on-site potentials $\epsilon_i$ with a uniform
distribution in the range $[-W/2, W/2]$. Hopping amplitude is set to
$t = 1$. The last term represents the chemical potential which
controls the particle density of the system. Disorder has no net
contribution to the chemical potential since we tune $\sum_{i=1}^{L}
\epsilon_i = 0$. Because the total number of particles conserves
in this system, we will restrict ourselves to the fixed total particle number at half-filling. Also zero temperature is assumed.

In the clean limit, up to a Jordan-Wigner transformation, this model is equivalent to the $S=1/2$ XXZ spin-chain model, for which there exists the Bethe-ansatz
solution that can be used to guide our numerical simulation.
For $|U| < 2$, the system is gapless (i.e. metallic);
For $U > 2$, the charge-density-wave (CDW) phase renders the system a Mott insulator;
For $U < -2$, the ground state becomes unstable to phase separation, being either empty or full filling.\cite{CNYang}
For the disordered case, an insulator to metal transition is predicted at $U \approx -1$;\cite{Rice_82PRB, Schulz_88PRB} A second metal to insulator transition happens at further stronger interaction, where the fermions start to form clusters and become localized again. In this Letter, we set our scope in the range $-2 < U < 0$.

The EE is numerically obtained with the density matrix renormalization group (DMRG) algorithm, which allows the calculation
of ground state and relevant physical quantities for large
interacting systems with high accuracy.\cite{White_DMRG} In our
case, we consider open wires of variant sizes up to
$L = 1024$. During the DMRG calculation, we retained 250 (for $L < 1024$) and
330 (for $L = 1024$) states for each block, and performed lattice sweeping until both the ground
state energy and the EE are converged.

\begin{figure}[tbph]
  \centering
   \includegraphics[width=0.51 \textwidth]{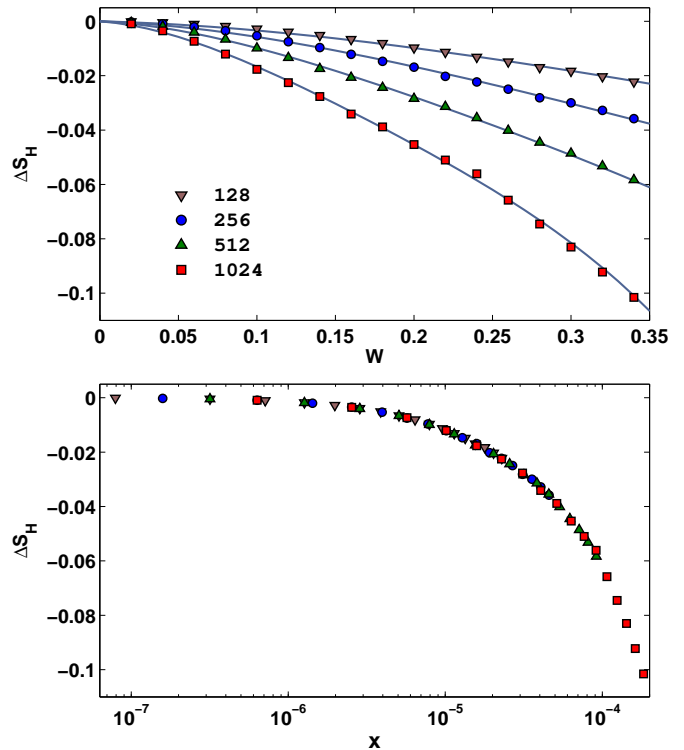}\\
  \caption{(color online). upper panel: scaling of $\Delta S_H$ for the non-interacting case against disorder strength with different system sizes, the solid lines represent the best fit of the scaling function with $n=2$; lower panel: data collapse to the one-parameter scaling function $f(x)$. }\label{figure1}
\end{figure}

 In the special case when $U=0$ the model recovers to a single particle system. In such case no true Anderson transition exists in the system as the only delocalized point is the clear limit where $W=0$. However, since it is well established that the localization length in such a system shows power-law divergence as $\xi \approx 105/W^2$ close to the clean limit (i.e. with a 'critical exponent' $\nu=2.0$),\cite{Romer_97PRL} we can use it as a benchmark for the scaling analysis. In this case the EE can be obtained quickly by the method developed by Peschel without the bother of DMRG.\cite{Peschel_03} Also averaging over a large number of random disorder configurations is allowed.

The scaling quantity adopted is $\Delta S_H=S_H -{S_H}'$, where $S_H$ and ${S_H}'$ are the EE of the clean and disordered systems respectively. The scaling behavior is shown in Fig.1 for the non-interacting case with system size up to $L=1024$. $10^4$ random disorder configurations have been collected for each data point of all system sizes. We then fit the dependence of $\Delta S_H$ on the system size $L$ and disorder $W$ to a one-parameter scaling law incorporated with a background term
\begin{equation}\label{dsh}
  \Delta S_H = f(x)+c(U,W),
\end{equation}
where the dimensionless $x$ is defined as $x=L/2\xi$. $c(U,W)$ is the background term which is scaling irrelevant.\cite{Slevin_QHE} Since $\xi$ is divergent in the critical regime, $x$ is vanishingly small. This allows us to do a power expansion of the scaling function
\begin{equation}
  f(x) =  x^{1/\nu} + a_2x^{2/\nu}+\ldots+a_{n}x^{n/\nu}. \label{scalingf}
\end{equation}
Here, $\nu$ is the critical exponent that characterizes the divergence of the localization length against the controlling parameter $\chi$
\begin{equation}
  \xi = \xi_0|\chi - \chi_c|^{-\nu}.  \label{xidiver}
\end{equation}
For the non-interacting case the only controlling parameter in our model is $W$ and obviously $W_c=0$. In Eq.\ref{scalingf}, $a_1$ has been set to unity to avoid redundancy in the fitting parameters, which also means the absolute value of $\xi_0$ is not determined in our fitting. Since in the non-interacting case when $x=0$ the system is exactly clean, which means $\Delta S_H=0$, and thus $c(0,0) = 0$.
We can also do a power expansion for the background term $c(0,W)$ against $W$. But during testing we find this term is vanishingly small, thus we will treat this term as zero for this very case.
This will reduce the total number of free fitting parameters to $n_t=n+1$. The fitting result is presented in Table. \ref{t1}. We find it is sufficient to terminate the expansion terms at $n=2$ for the scaling function. Considering more expansion terms yields consistent results. We have also tested the fitting stability using different ranges of $W$, the result also shows good consistency.(see Table.\ref{t3}) The estimated 'critical exponent' $\nu$ agrees precisely with that given by the well-established transfer-matrix method in Ref.\cite{Romer_97PRL}.
\begin{table}
\caption{Best fit of the critical exponent for the non-interacting case with different number of expansion terms. $0<W<0.35$.}\label{t1}
\centering\begin{tabular}{p{1.5cm} p{1.5cm} p{1.5cm} p{1.5cm}}
\hline\hline\noalign{\smallskip}
   n & 2 & 3 & 4    \\
\hline\noalign{\smallskip}
   $\nu$  & 2.04   & 2.03 & 1.99    \\
\hline\hline
\end{tabular}

\end{table}
\begin{table}
\caption{Same with Table.\ref{t1} but fixing $n=4$ and using different ranges of $W$.}\label{t3}
\centering\begin{tabular}{p{1.5cm} p{1.5cm} p{1.5cm} p{1.5cm}}
\hline\hline\noalign{\smallskip}
   W & (0, 0.35) & (0, 0.25) & (0, 0.15)    \\
\hline\noalign{\smallskip}
   $\nu$  & 1.99   & 1.99 & 2.00    \\
\hline\hline
\end{tabular}
\end{table}

To verify the validity of the one-parameter scaling of the EE, we need to see how the data collapses to the scaling function, which supposedly is a universal fit to the data of all systems regardless of the size $L$. This is shown in the lower panel of Fig.1, where the data from the upper panel is re-plotted as a function of the dimensionless $x$. All points collapses nicely to a single curve which corresponds to a fit with $n=2$. Although the absolute value of $\xi_0$ (and hence $\xi$) is not determined in the fitting, it does not affect the data collapse since what $\xi_0$ does is merely shifting the curve horizontally in the plot.

\begin{figure}[tbhp]
  \centering
  \includegraphics[width=0.51 \textwidth]{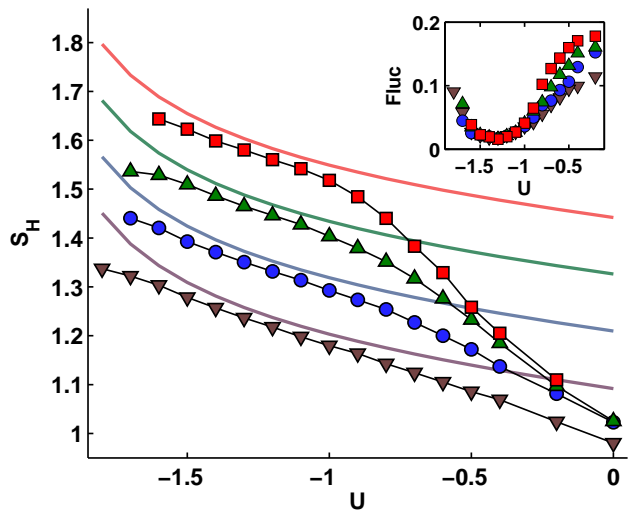}\\
  \caption{(color online).  $S_H$ for the interacting case of both clean (solid lines) and corresponding disordered (dot lines) systems for different system sizes. The inset shows the corresponding fluctualtions of $S_H$. In the delocalized phase, the EE's fluctuation is found to be minimum. $W=1.0$.}\label{figure2}
\end{figure}

Next we switch the interaction on and calculate the EE with the DMRG algorithm. In this case our aim is to identify the interaction induced Anderson localization-delocalization transition. We choose a fixed disorder strength $W=1.0$ and sweep $U$. For the disordered systems the EE is averaged over 1000 (for $L < 1024$) and 400 (for $L = 1024$) random disorder configurations.

The EE of both clean (solid lines) and disordered (dotted lines) systems are plotted in Fig. 2. For the clean case the EE scales as $\frac{1}{6}\mathrm{log} L$. However, for the disordered case the EE is restricted by localization and hence deviates from logarithmic scaling. But in a certain range of $U$ the EE recovers to the same logarithmic scaling as the clean case. In the weak interaction regime, the EE's difference between the clean and disordered case is clearly dependent on both the system size and the interaction. When the interaction becomes stronger this dependence gradually disappears but finally re-emerges. The same trend can also be observed in the fluctuations of the EE.(inset of Fig. 2) This signatures two transitions: first from localization to delocalization and then from delocalization to localization. In the delocalized phase $\Delta S_H$ is only allowed to be a scaling irrelevant quantity since both $S_H$ and ${S_H}'$ scales as $\frac{1}{6}\mathrm{log} L$ in this regime. In Fig.2, this regime locates roughly at $-1.5 < U < 1$. The same observation can be made when fixing $U$ and sweep $W$. As is shown in Fig.4, in a certain range of $W$ the $\Delta S_H$ becomes size independent but is a function of $W$ only. This phenomena itself can be used to give a rough estimate of the delocalized phase, just as is done in Ref.\cite{PS_98PRL} using the ground state phase sensitivity.

\begin{figure}[tbph]
  \centering
  \includegraphics[width=0.51 \textwidth]{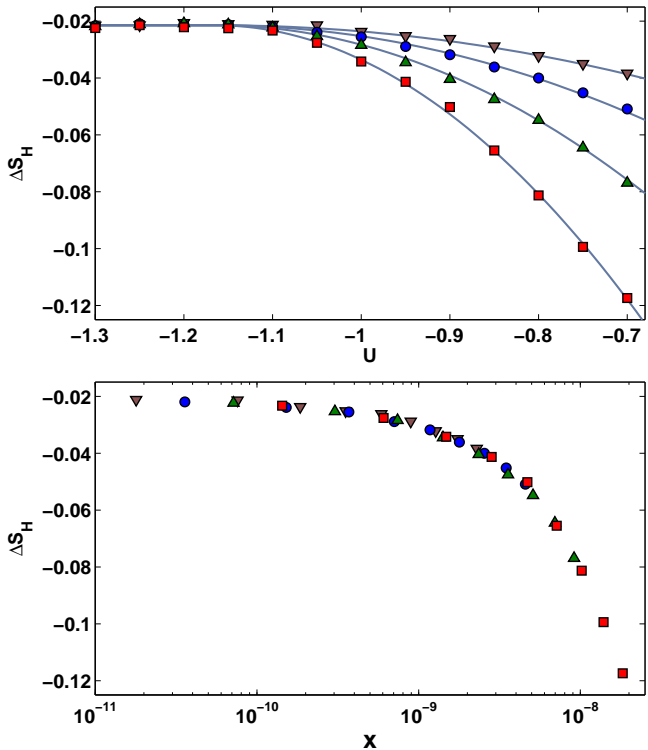}\\
  \caption{(color online). upper panel: scaling of $\Delta S_H$ against interaction with different system sizes, the solid lines represent the best fit of the scaling function with n=2; lower panel: data collapse to the one-parameter scaling function $f(x)$. $W=1.0$.}\label{figure4}
\end{figure}

 To qualitatively characterize the phase transition, we need more extensive simulations at the vicinity of the transition point. We will fix $W$ and use $U$ as the controlling parameter in the FSS. Due to the system's instability to phase separation at strong interactions, which is exaggerated by the disorder, we find it impossible to stick to half-filling in the DMRG calculations when $U < -1.8$. We thus choose to demonstrate the FSS analysis at the first Anderson transition where sufficient reliable EE data can be collected.

 \begin{table}
\caption{Best fit of the critical parameters for the interacting case with different number of expansion terms. $W=1.0.$}\label{t2}
\centering\begin{tabular}{p{1.5cm} p{1.5cm} p{1.5cm} p{1.5cm}}
\hline\hline\noalign{\smallskip}
   n & 2 & 3 & 4    \\
\hline\noalign{\smallskip}
   $\nu$  & 2.33   & 2.31 & 2.35    \\
   \hline\noalign{\smallskip}
   $U_c$  & -1.15   & -1.15 & -1.15    \\
\hline\hline
\end{tabular}
\end{table}

\begin{figure}[tbph]
  \centering
  \includegraphics[width=0.51 \textwidth]{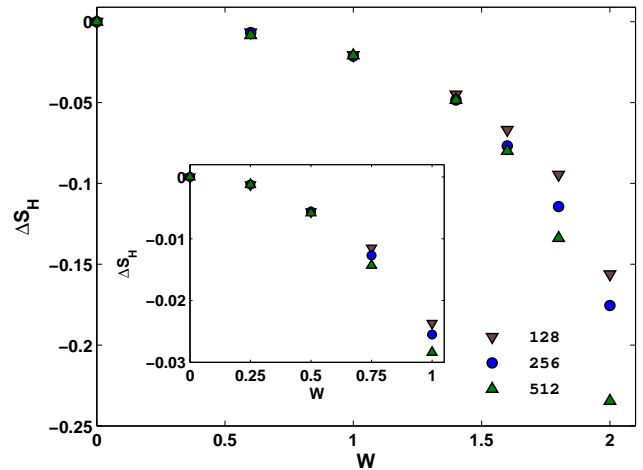}\\
  \caption{(color online). $\Delta S_H$ for different system sizes as a function of $W$ when fixing $U = -1.2$; inset: $U=-1.0$.}\label{figure3}
\end{figure}

In the upper panel of Fig. 3 we plot the scaling behavior of $\Delta S_H$ against $U$. The delocalized phase clearly appears as a plateau in the plot. $\Delta S_H$ shows almost no dependence on $U$ in the delocalized phase as is compared with Fig. 4 in which it clearly is a function of $W$. This allows us to treat the background term $c(U,1.0)$ as a small constant $a_0$ in the scaling function Eq. \ref{dsh}. We will assume $\xi=\infty$ when $U<U_c$. Accordingly Eq. \ref{xidiver} is modified as
 \begin{equation}
   \xi^{-1}=\theta(U-U_c){\xi_0}^{-1}(U - U_c)^\nu
 \end{equation}
 in the fitting, where $\theta(x)$ is the step function. In this case the total number of free fitting parameters is $n_t=n+3$ since both $a_0$ and $U_c$ also needs to be determined by fitting. We have also considered different number of expansion terms for the scaling function. The result shows good consistency when using $n\geq2$. As is listed in Table.\ref{t2}, when increasing $n$ the estimated $U_c$ shows almost no change, only $\nu$ slightly varies. In the lower panel of Fig. 3, the one-parameter scaling is demonstrated graphically by re-plotting the data with $U>U_c$ as a function of the dimensionless $x$. Again the absolute value of $\xi$ is not determined but only the critical exponent $\nu$ that describes $\xi$'s divergence is meaningful. The obtained $U_c$ agrees with the roughly estimated phase boundary of Ref.\cite{PS_98PRL}. But both $U_c$ and $\nu$  deviates from Ref.\cite{Cater_05PRB}, in which no FSS is involved and the estimated $\xi$ is limited to the maximum system size $L$ instead of diverging.

In summery, we have shown that the EE is an efficient quantity in characterizing the Anderson transition in one dimension. The demonstrated FSS of the EE can be used to quantitatively characterize the the Anderson transitions in 1D systems. The precision of the method can be improved by collecting a large number of random disorder configurations for averaging. This is particularly feasible for the non-interacting systems where calculation of EE is not computationally demanding. This is evident by comparing the data collapse in our Fig. 1 and Fig. 3. Obviously the more intensive sample averaging in Fig. 1 has helped to produce better data collapse and more precise determinant of the scaling function. For the interacting systems, since the calculation of the EE is naturally incorporated with the standard DMRG algorithm, the method can also be readily applied.

We gratefully acknowledge discussions with Cenke Xu and Chuanwei Zhang. This work is supported by Research Grant Council of Hong Kong under Grant No. HKU 705110P. Chu is also supported by AFOSR (FA9550-11-1-0313) and ARO (W911NF-09-1-0248).


\begin{references}
\bibitem{Anderson}P. W. Anderson, Phys. Rev. 109, 1492 (1958).
\bibitem{AALR}E. Abrahams, P. W. Anderson, D. C. Licciardello, and T. V. Ramakrishnan, Phys. Rev. Lett. 42, 673 (1979).
\bibitem{PLee_RMP}P. A. Lee and T. V. Ramakrishnan, Rev. Mod. Phys. 57, 287 (1985).

\bibitem{Rice_82PRB} W. Apel and T. M. Rice, Phys. Rev. B, 26, 7063 (1982). 
\bibitem{Schulz_88PRB}T. Giamarchi and H. J. Schulz, Phys. Rev. B 37, 325  (1988).
\bibitem{Slevin_QHE} K. Slevin and T. Ohtsuki, Phys. Rev. B 80, 041304(R) (2009).
\bibitem{PS_98PRL} P. Schmitteckert, T. Schulze, C. Schuster, P. Schwab, and U. Eckern, Phys. Rev. Lett., 80, 560 (1998); P.
Schmitteckert, R. A. Jalabert, D. Weinmann, and J. L. Pichard, Phys. Rev. Lett.81, 2308 (1998). 
\bibitem{Schuster_02PRB}C. Schuster, R. A. R$\ddot{o}$mer, and M. Schreiber, Phys. Rev. B 65,115114 (2002).
\bibitem{Cater_05PRB}J. M. Carter and A. MacKinnon, Phys. Rev. B, 72, 024208 (2005). 
\bibitem{MacKinnon}A. MacKinnon and B. Kramer, Z. Phys. B 53, 1 (1983).
\bibitem{Huckestein} B. Huckestein, Rev. Mod. Phys. 67, 357 (1995).

\bibitem{Slevin_QSH}Y. Asada and K. Slevin, T. Ohtsuki, Phys. Rev. Lett. 89, 256601 (2002).


\bibitem{EE_RMP}For a review see: L. Amico, R. Fazio, A. Osterloh, and V. Vedral, Rev. Mod. Phys. 80, 517 (2008).
\bibitem{Vidal_03PRL}G. Vidal, J. I. Latorre, E. Rico, and A. Kitaev, Phys. Rev. Lett, 90, 227902 (2003).
\bibitem{Cardy_JSM}P. Calabrese and J. Cardy, J. Stat. Mech. (2004) P06002.

\bibitem{Ryu_06PRB}S. Ryu and Y. Hatsugai, Phys. Rev. B 73, 245115 (2006).
\bibitem{Refael_04PRL}G. Refael and J. E. Moore, Phys. Rev. Lett.,93, 260602 (2004). 
\bibitem{Pollman_10NJP}F. Pollmann and J. E. Moore, New J. Phys. 12, 025006. (2010).
\bibitem{Turner_11PRB}A. M. Turner, F. Pollmann and E. Berg, Phys. Rev. B 83, 075102 (2011).
\bibitem{Luca} L. Tagliacozzo, T. R. de Oliveira, S. Iblisdir, and J. I. Latorre Phys. Rev. B 78, 024410 (2008).

\bibitem{Fagotti_11PRB}M. Fagotti, P. Calabrese, and J. E. Moore, Phys. Rev. B 83, 045110 (2011).
\bibitem{Igloi_08JSM}F. Igloi and Y. C. Lin, J. Stat. Mech. (2008) P06004.
\bibitem{Jia_08PRB} X. Jia, A. R. Subramaniam, I. A. Gruzberg, and S. Chakravarty, Phys. Rev. B 77, 014208 (2008). 
\bibitem{Bardarson_12PRL}J. H. Bardarson, F. Pollmann, and J. E. Moore Phys. Rev. Lett., 109, 017202 (2012).


\bibitem{Laflorencie} N. Laflorencie, Phys. Rev. B 72 140408 (2005).
\bibitem{Santachiara} R. Santachiara, J. Stat Mech, L06002 (2006).  

\bibitem{Berkovits_12PRL}R. Berkovits, Phys. Rev. Lett., 108,176803 (2012). 
\bibitem{CNYang}C. N. Yang and C. P. Yang, Phys. Rev. 150, 321; Phys. Rev. 150, 327 (1966).
\bibitem{White_DMRG}S. R. White, Phys. Rev. Lett. 69, 2863 (1992); S. R. White
and R. M. Noack, Phys. Rev. Lett. 68, 3487 (1992); S. R. White, Phys. Rev. B 48, 10 345 (1993).
\bibitem{Pollmann_09PRL}F. Pollmann, S. Mukerjee, A. M. Turner, and J. E. Moore, Phys. Rev. Lett., 102, 255701 (2009).
\bibitem{Romer_97PRL} R. A. R$\ddot{o}$mer and M. Schreiber, Phys. Rev. Lett. 78, 515 (1997).
\bibitem{Peschel_03}I. Peschel, J. Phys. A: Math. Gen. 36 L205 (2003).


\end{references}
\end{document}